\documentclass[runningheads,11pt,times,psfig]{llncs}
\usepackage{makeidx}  
\usepackage[T1]{fontenc}
\usepackage[latin1]{inputenc}
\usepackage{epic}
\usepackage{eepic}
\usepackage{alltt}
\usepackage{graphicx}
\usepackage{amssymb,amsmath,algorithm}
\input{psfig.sty}

\newcommand{\tb}{\makebox[0.4cm]{}}
\newcommand{\due}{\makebox[0.8cm]{}}
\newcommand{\tre}{\makebox[1.2cm]{}}

\def\squarebox#1{\hbox to #1{\hfill\vbox to #1{\vfill}}}

\newcommand{\ignore}[1]{}
\newcommand{\save}[1]{}

\newcommand\pulse{\mbox{\slshape pulse}}
\newcommand\bsigma{\bar\sigma}
\newcommand\cycle{\mbox{\slshape cycle}}
\newcommand\Cycle{\mbox{\slshape Cycle}}

\newcommand\cyclemin{\cycle_{\mbox{\scriptsize\slshape min}}}
\newcommand\cyclemax{\cycle_{\mbox{\scriptsize\slshape max}}}
\newcommand\ByzConsensus{\mbox{\sc Byz\_Consensus}}
\newcommand\ApproxByzAgreement{\mbox{\sc Approx\_Byz\_Agree}}
\newcommand\ByzAgreement{\mbox{\sc Byz\_Agreement}}
\newcommand\ClockSynch{\mbox{\sc PBSS-Clock-Synch}}
\newcommand\CYCLECS{\mbox{\sc Cycle-Wrap-CS\ }}
\newcommand\ApproxClocks{\mbox{\sc Approx-CS}}
\newcommand\ConsBCast{\mbox{\sc Consensus-broadcast\ }}
\newcommand\Broadcast{\mbox{\sc Broadcast\ }}

\begin{document}

\pagestyle{headings}  

\mainmatter
\title{Linear-time Self-stabilizing Byzantine Clock Synchronization\\(\small updated version)}
\titlerunning{Self-stabilizing Byzantine Clock Synchronization}

\author{ Ariel Daliot\inst{1} \and Danny
Dolev\inst{1} \and Hanna Parnas\inst{2}}
\authorrunning{Daliot, Dolev and Parnas}

\institute{School of Engineering and Computer Science, The Hebrew
University of Jerusalem, Israel.
\email{\{adaliot,dolev\}@cs.huji.ac.il} \and Department of
Neurobiology and the Otto Loewi Minerva Center for Cellular and
Molecular Neurobiology, Institute of Life Science, The Hebrew
University of Jerusalem, Israel.  \email{hanna@vms.huji.ac.il} }

\maketitle

\ignore{
\title{Linear-time Self-stabilizing Byzantine Clock Synchronization}

\author{Ariel Daliot, Danny Dolev and Hanna Parnas}

\institute{School of Engineering and Computer Science, The Hebrew
University of Jerusalem, Israel, \{dolev,adaliot\}@cs.huji.ac.il}

 \maketitle
}

\begin{abstract} Clock synchronization is a very
fundamental task in distributed system. The vast majority of
distributed tasks require some sort of synchronization and clock
synchronization is a very straightforward tool for supplying this.
It thus makes sense to require an underlying clock synchronization
mechanism to be highly fault-tolerant. A self-stabilizing algorithm
seeks to attain synchronization once lost; a Byzantine algorithm
assumes synchronization is never lost and focuses on containing the
influence of the permanent presence of faulty nodes. There are
efficient self-stabilizing solutions for clock synchronization as
well as efficient solutions that are resilient to Byzantine faults.
In contrast, to the best of our knowledge there is no practical
solution that is self-stabilizing while tolerating the permanent
presence of Byzantine nodes. Designing algorithms that
self-stabilize while at the same time tolerate permanent Byzantine
failures present a special challenge due to the ``ambition'' of
malicious nodes to hamper stabilization if the system tries to
recover from a corrupted state. We present the first linear-time
self-stabilizing Byzantine clock synchronization algorithm. Our
deterministic clock synchronization algorithm is based on the
observation that all clock synchronization algorithms require events
for exchanging clock values and re-synchronizing the clocks to
within safe bounds. These events usually need to happen
synchronously at the different nodes. In classic Byzantine
algorithms this is fulfilled or aided by having the clocks initially
close to each other and thus the actual clock values can be used for
synchronizing the events. This implies that clock values cannot
differ arbitrarily, which necessarily renders these solutions to be
non-stabilizing. Our scheme suggests using an underlying distributed
pulse synchronization module that is uncorrelated to the clock
values. The synchronized pulses are used as the events for
re-synchronizing the clock values. The algorithm is very efficient
and attains and maintains high precision of the clocks.
\end{abstract}

{\em This is an updated version. The original paper appeared in
OPODIS'03. The main difference is the replacement of the pulse
synchronization module.}

\ignore{ \textbf{Keywords:} Clock synchronization, Pulse
synchronization, Self-stabilization, Internal clock
synchronization, Byzantine faults}

\section{Introduction}
\label{sec:intro}

On-going faults whose nature is not predictable or that express
complex behavior are most suitably addressed in the Byzantine fault
model. It is the preferred fault model in order to seal off
unexpected behavior within limitations on the number of concurrent
faults. Most distributed tasks require the number of concurrent
Byzantine faults, $f,$ to abide by the ratio of $3f<n,$ where $n$ is
the network size. See~\cite{Impossibility86} for impossibility
results on several consensus related problems such as clock
synchronization. Additionally, it makes sense to require systems to
resume operation after a major failure without the need for an
outside intervention and/or a restart of the system from scratch.
E.g. systems may occasionally experience short periods in which more
than a third of the nodes are faulty or messages sent by all nodes
may be lost for some time due to a network failure.

Such transient violations of the basic fault assumptions may leave
the system in an arbitrary state from which the protocol is required
to resume in realizing its task. Typically, Byzantine algorithms do
not ensure convergence in such cases, as strong assumptions are
usually made on the initial state and thus merely focus on
preventing Byzantine faults from notably shifting the system state
away from the goal. A \textit{self-stabilizing} algorithm bypasses
this limitation by being designed to converge within finite time to
a desired state from any initial state. Thus, even if the system
loses its consistency due to a transient violation of the basic
fault assumptions (e.g. more than a third of the nodes being faulty,
network disconnected, etc.), then once the system becomes coherent
again the protocol will successfully realize the task, irrespective
of the resumed state of the system. In trying to combine both fault
models, Byzantine failures present a special challenge for designing
stabilizing algorithms due to the ``ambition'' of malicious nodes to
incessantly hamper stabilization, as might be indicated by the
remarkably few algorithms resilient to both fault models. For a
short survey of self-stabilization see~\cite{R27}, for an extensive
study see~\cite{R22}.

The current paper addresses the problem of synchronizing clocks in a
distributed system. There are several efficient algorithms for
self-stabilizing clock synchronization withstanding crash faults
(see~\cite{DW97b,PT97,DPossImp97}, for other variants of the problem
see~\cite{ADG91,H00b}). There are many efficient classic Byzantine
clock synchronization algorithms, for a performance evaluation of
clock synchronization algorithms see~\cite{CSEVAL98}. However,
strong assumptions on the initial state of the nodes are typically
made, usually assuming all clocks are initially synchronized
(\cite{CSEVAL98,R3,WelchLynch88}) and thus these are not
self-stabilizing solutions. On the other hand, self-stabilizing
clock synchronization algorithms allow initialization with arbitrary
clock values, but typically have a cost in the convergence times or
in the severity of the faults contained. Evidently, there are very
few self-stabilizing solutions facing Byzantine faults
(\cite{DolWelSSBYZCS04}), all with unpractical convergence times.
The protocols in \cite{DolWelSSBYZCS04} are to the best of our
knowledge the first self-stabilizing protocols that are tolerant to
Byzantine faults. Note that self-stabilizing clock synchronization
has an inherent difficulty in estimating real-time without an
external time reference due to the fact that non-faulty nodes may
initialize with arbitrary clock values. Thus, self-stabilizing clock
synchronization aims at reaching a legal state from which clocks
proceed synchronously at the rate of real-time (assuming that nodes
have access to physical timers which rate is close to real-time) and
not necessarily at estimating real-time. Many applications utilizing
the synchronization of clocks do not really require the exact
real-time notion (see \cite{Liskov:1991}). In such applications,
agreeing on a common clock reading is sufficient as long as the
clocks progress within a linear envelope of any real-time interval.

We present a Byzantine self-stabilizing clock synchronization
protocol with the following property: should the system initialize
or recover from any transient faults with arbitrary clock values
then the clocks of the correct nodes proceed synchronously at
real-time rate. Should the clocks of the correct nodes hold values
that are close to real-time, then the correct clocks proceed
synchronously with high real-time accuracy. Thus, the protocol we
present significantly improves upon existing Byzantine
self-stabilizing clock synchronization algorithms by reducing the
time complexity from expected exponential (\cite{DolWelSSBYZCS04})
to deterministic $O(f).$ Our protocol improves upon existing
Byzantine non-stabilizing clock synchronization algorithms by
providing self-stabilization while performing with similar
complexity. The self-stabilization and comparably low complexity is
achieved by executing on top of a deterministic Byzantine
self-stabilizing algorithm for pulse synchronization
\cite{NEW-PULSE-TR}. The interval between the synchronized pulses is
long enough to allow initialization and termination of a Byzantine
consensus procedure on the clock values, thus attaining and
maintaining a common clock reading.

Having access to an outside source of real-time is useful. In such
case our approach maintains a consistent system state when the
outside source fails.

A special challenge in self-stabilizing clock synchronization is
the clock wrap around. In non-stabilizing algorithms having a
large enough integer eliminates the problem for any practical
concern. In self-stabilizing schemes a transient failure can cause
clocks to hold arbitrary large values, surfacing the issue of
clock bounds. Our clock synchronization scheme handles clock wrap
around difficulties.

The system may be in an arbitrary state in which the communication
network may behave arbitrarily and in which there may be an
unbounded number of concurrent Byzantine faulty nodes. The algorithm
will eventually converge once the communication network resumes
delivering messages within bounded, some $d,$ time units, and the
fraction of Byzantine nodes, $f,$ obeys $n \ge 3f+1,$ for a network
of size $n.$ The attained clock precision and accuracy is $11d$
real-time units, though we present an additional scheme that can
attain clock precision and accuracy of $3d.$ The convergence time is
$O(f^\prime)$ communication rounds, where $f^\prime \le f$ is the
actual number of concurrent faults. Our protocol has the additional
advantage of a minimal time and message overhead during steady-state
after the clocks have synchronized.

An additional advantage of our algorithm is the use of a Byzantine
Consensus protocol that works in a message driven manner. The basic
protocol follows closely the early stopping Byzantine Agreement
protocol of Toueg, Perry and Srikanth~\cite{FastAgree87}. The main
difference is that the protocol rounds progress at the rate of the
actual time of information exchange among the correctly operating
nodes. This, typically, is much faster than progression with rounds
whose time lengths are functions of the upper bound on message
delivery time between correct nodes.

\section{Model and Problem Definition}
\label{sec:model}

The environment is a bounded-delay network model of $n$ nodes that
communicate by exchanging messages. We assume that the message
passing allows for an authenticated identity of the senders. The
communication network does not guarantee any order on messages among
different nodes. Individual nodes have no access to a central clock
and there is no external pulse system. The hardware clock rate
(referred to as the {\em physical timers}) of correct nodes has a
bounded drift, $\rho,$ from real-time rate. Consequent to transient
failures there can be an arbitrary number of concurrent Byzantine
faulty nodes, the turnover rate between faulty and non-faulty
behavior of the nodes can be arbitrary and the communication network
may behave arbitrarily. Eventually the system behaves coherently
again but in an arbitrary state.

\begin{definition} A node is {\bf non-faulty} at times that it complies with the
following:
\begin{enumerate}
\vspace{-0.5em} \item Obeys a global constant $0<\rho<<1$ (typically
$\rho \approx 10^{-6}$), such that for every real-time interval
$[u,v]:$\vspace{-2mm}
$$(1-\rho)(v-u)  \le \mbox{ `physical timer'}(v) -
\mbox{ `physical timer'}(u) \le (1+\rho)(v-u).$$\vspace{-6mm}

\item Operates according to the instructed protocol.

\item Processes any message of the instructed protocol within $\pi$ real-time units of arrival time.

\end{enumerate}
\end{definition}

A node is considered {\bf faulty} if it violates any of the above
conditions. We allow for Byzantine behavior of the faulty nodes. A
faulty node may recover from its faulty behavior once it resumes
obeying the conditions of a non-faulty node. For consistency
reasons, the ``correction'' is not immediate but rather takes a
certain amount of time during which the non-faulty node is still not
counted as a correct node, although it supposedly behaves
``correctly''\footnote{For example, a node may recover with
arbitrary variables, which may violate the validity condition if
considered correct prematurely.}. We later specify the time-length
of continuous non-faulty behavior required of a recovering node to
be considered \textbf{correct}.

\begin{definition} The communication network is {\bf non-faulty} at periods that it complies with the
following: \label{def:net_nonfaulty}
\begin{enumerate}

\item Any message sent by any non-faulty node
arrives at every non-faulty node within $\delta$ real-time units;

\item All messages sent by a non-faulty node and received by a non-faulty node obey FOFI order.

\end{enumerate}
\end{definition}

 The system is said to be coherent only following some
 minimal\footnote{An infinitely small time period
in which the nodes and the communication network are non-faulty has
no practical meaning. The required minimal value in our context will
be specified later.} amount of time of continuous non-faulty
behavior of the nodes and the communication network.

\vspace{+2mm} \noindent{\bf Basic notations: }

We use the following notations though nodes do not need to
maintain all of them as variables.
\begin{itemize}
\vspace{-0.5em}

\item $d\equiv \delta + \pi.$ Thus, when the communication network is non-faulty,
$d$ is the upper bound on the elapsed real-time from the sending of
a message by a non-faulty node until it is received and processed by
every correct node.

\item $Clock_i,$ the clock of node $i,$ is a real value in the
range $0$ to $M-1.$ Thus $M-1$ is the maximal value a clock can
hold. Its progression rate is a function of node $p_i$'s physical
timer. The clock is incremented every time unit. $Clock_i(t)$
denotes the value of the clock of node $p_i$ at real-time $t.$

\item $\gamma$ is the target upper bound on the difference of
clock readings of any two correct clocks at any real-time. Our
protocol achieves $\gamma = 3d+O(\rho).$

\item Let $a, b, g, h \in R^+$ be constants that define the linear
envelope bound of the correct clock progression rate during any
real-time interval.

\item $\Psi_i(t_1,t_2)$ is the amount of clock time elapsed on the
clock of node $p_i$ during a real-time interval $[t_1,t_2]$ within
which $p_i$ was continuously correct. The value of $\Psi$ is not
affected by any wrap around of $clock_i$ during that period.

\item A \textbf{$\pulse$} is an internal event targeted to happen in tight synchrony at all correct nodes.
A \textbf{$\Cycle$} (with upper-case initial letter) is the
``ideal'' time interval length between two successive pulses that a
node invokes, as given by the user. The actual cycle length, denoted
with lowercase initial, has upper and lower bounds as a result of
faulty nodes and the physical clock skew, denoted $\cyclemax$ and
$\cyclemin$ respectively.

\item $\sigma$ represents the upper bound on the real-time between
the invocation of the pulses of different correct nodes ({\em
tightness of pulse synchronization}). The pulse synchronization
procedure in \cite{NEW-PULSE-TR} achieves $\sigma=3d.$

\item $\mbox{\slshape pulse\_conv}$ represents the
convergence time of the underlying pulse synchronization module. The
pulse procedure in \cite{NEW-PULSE-TR} converges within
$6\cdot\cycle.$

\item $\mbox{\slshape agreement\_duration}$ represents the maximum
real-time required to complete the chosen Byzantine consensus
procedure used in
Section~\ref{sec:basic}. We assume\\
$\sigma\le\sigma+\mbox{\slshape agreement\_duration}< \cycle \le
\Cycle + \mbox{\slshape agreement\_duration}.$ For simplicity of our
arguments we also assume that $M>\mbox{\slshape
agreement\_duration}$ but this is not a necessary assumption.
\end{itemize}

Non-faulty nodes do not initialize with arbitrary values of $n,$ $f$
and $\Cycle$ as these are fixed constants. It is required that
$\Cycle$ is chosen s.t. $\cyclemin$ is large enough to allow our
protocol to terminate in between pulses.

A recovering node should be considered correct only once it has been
continuously non-faulty for enough time to enable it to go through a
complete ``synchronization process''. This is the time it takes,
from any state, to complete a pulses that is in synchrony with all
other correct nodes and synchronize with the consensus variables.

\begin{definition} The communication network is {\bf correct}
following $\Delta_{net}$ real-time of continuous non-faulty
behavior.\footnote{We will use $\Delta_{net}\ge \mbox{\slshape
pulse\_conv} +\mbox{\slshape agreement\_duration}+\sigma.$}
\end{definition}

\begin{definition} A node is {\bf correct} following $\Delta_{node}$  real-time of continuous
non-faulty behavior during a period that the communication network
is correct.\footnote{We will use $\Delta_{node}\ge \mbox{\slshape
pulse\_conv} +\mbox{\slshape agreement\_duration}+\sigma.$}
\end{definition}

\begin{definition}\label{def:system-coherence} The system is said to
be {\bf coherent} at times that it complies with the following:

\begin{enumerate}

\item \emph{(Quorum)} At least $n-f$ of the nodes are correct, where $n\ge 3f+1;$

\item \emph{(Network Correctness)} The communication network is correct.
\end{enumerate}
\end{definition}

Hence, if the system is not coherent then there can be an unbounded
number of concurrent faulty nodes; the turnover rate between faulty
and non-faulty nodes can be arbitrarily large and the communication
network may behave arbitrarily. When the system is coherent, then
the communication network and a large enough fraction of the nodes
($n-f$) have been non-faulty for a sufficiently long time period for
the pre-conditions for convergence of the protocol to hold. The
assumption in this paper, as underlies any other self-stabilizing
algorithm, is that eventually the system becomes coherent.

\vspace{+2mm} \noindent{\bf Basic definitions:} \vspace{-2mm}
\begin{itemize}
\item The {\bf clock\_state} of the system at real-time $t$ is
given by: \vspace{-2mm} $$clock\_state(t) \equiv (clock_0(t),\;
\ldots,\; clock_{n-1}(t))\;.$$

\vspace{-2mm}

\item The systems is in a {\bf synchronized clock\_state} at
real-time $t$ if $\forall\, correct\  p_i, p_j ,$\vspace{-2mm}
$$(|clock_i(t)-clock_j(t)|  \le \gamma) \ \vee \
(|clock_i(t)-clock_j(t)|  \ge M - \gamma)\;.\footnote{The second
condition is a result of dealing with bounded clock variables.}$$
\end{itemize}\vspace{-2mm}

\begin{definition} {\bf The ``Self-stabilizing Byzantine Clock Synchronization Problem''}

\noindent \begin{description}

\item{\bf Convergence:} Starting from an arbitrary system state,
$s,$ the system reaches a synchronized clock\_state after a finite
time.
 \vspace{+2mm}
\item{ {\bf Closure:}} If $s$ is a synchronized clock\_state of
the system at real-time $t_0$ then $\forall\, real\ time\ t\ge
t_0,$
\begin{enumerate}
\item clock\_state(t) is a synchronized clock\_state,
 \item ``Linear Envelope'': for every correct
 node, $p_i,$\\
 \vspace{-2mm} $$a\cdot[t-t_0] + b   \le
 \Psi_i(t_0,t)
 \le  g\cdot[t-t_0] + h\;.$$ \vspace{-5mm}
 \end{enumerate}

\end{description}
\end{definition}

The second Closure condition intends to bound the effective clock
progression rate in order to defy a trivial solution.

\section{Self-stabilizing Byzantine Clock Synchronization}
\label{sec:clock-synch} \vspace{-2mm} A major challenge of
self-stabilizing clock synchronization is to ensure clock
synchronization even when nodes may initialize with arbitrary clock
values. This, as mentioned before, requires handling the wrap around
of clock values. The algorithm we present employs as a building
block an underlying self-stabilizing Byzantine pulse synchronization
procedure presented in \cite{NEW-PULSE-TR}. In the pulse
synchronization problem nodes invoke pulses regularly, ideally every
$\Cycle$ time units. The goal is for the different correct nodes to
do so in tight synchrony of each other. To synchronize their clocks,
nodes execute at every pulse Byzantine consensus on the clock value
to be associated with the next pulse event\footnote{It is assumed
that the time between successive pulses is sufficient for a
Byzantine consensus algorithm to initiate and terminate in
between.}. When pulses are synchronized, then the consensus results
in synchronized clocks. The basic algorithm uses strong consensus to
ensure that once correct clocks are synchronized at a certain pulse,
and thus enter the consensus procedure with identical values, then
they terminate with the same identical values and keep the
progression of clocks continuous and synchronized\footnote{The pulse
synchronization building block does not use the value of the clock
to determine its progress, but rather intervals measured on the
physical timer.}.

\subsection{The Basic Clock Synchronization Algorithm}
\label{sec:basic} \vspace{-2mm} The basic clock synchronization
algorithm is essentially a self-stabilizing version of the Byzantine
clock synchronization algorithm in~\cite{R3}.

We call it \ClockSynch\ (for \textit{Pulse-based Byzantine
Self-stabilizing Clock Synchronization}). The agreed clock time to
be associated with the next pulse (next ``time for synchronization''
in~\cite{R3}) is denoted by ET (for \textit{Expected Time}, as
in~\cite{R3}). Synchronization of clocks is targeted to happen every
$\Cycle$ time units, unless the pulse is invoked earlier (or
later)\footnote{$\Cycle$ has the same function as PER
in~\cite{R3}.}.

\ignore{\vspace{2mm}\noindent
{\bf \ClockSynch}\\
{\bf at} ``$pulse$'' event \hfill\textit{/* received the internal pulse event */}\\
\tb {\bf begin} \\
\due        $Clock      := ET;$\\
\due        Abort possible running instance of \ClockSynch\ and Reset all buffers;\\
\due        Wait $\sigma(1+ \rho)$ time units; \\
\due        $Next\_ET$  := \ByzConsensus($(ET + \Cycle)\bmod M,\;\sigma$);\\
\due        $Clock      := (Clock + Next\_ET-(ET+\Cycle)) \bmod M;$ \hfill\textit{/* posterior adj.}\\
\due        $ET         := Next\_ET;$\\
\tb {\bf end}\\

\begin{figure}[!h]
\center \fbox{\begin{minipage}{4.5in} \footnotesize
\setlength{\baselineskip}{3.5mm}

{\bf \ClockSynch}\\
{\bf at} ``$\pulse$'' event \hfill\textit{/* received the internal pulse event */}\\
{\bf begin} \\
\tb        $Clock      := ET;$\\
\tb        $Timer := 0$;\\
\tb        Abort any other running instance of \ClockSynch\ \\
\tre\tre and procedures it has invoked;\\
\tb        Wait until $Timer=\sigma(1+ \rho)$ time units; \\
\tb        $Next\_ET$  := \ByzConsensus($(ET + \Cycle)\bmod M,\;\sigma$);\\
\tb        $Clock      := (Clock + Next\_ET-(ET+\Cycle)) \bmod M;$ \hfill\textit{/* posterior adj.}\\
\tb        $ET         := Next\_ET;$\\
{\bf end}
\normalsize
\end{minipage} }
\caption{The self-stabilizing Byzantine clock synchronization
protocol} \label{alg:pbss}
\end{figure}}

\begin{figure}[!h]
\center \fbox{\begin{minipage}{4.92in} \footnotesize
\setlength{\baselineskip}{3.5mm}

Algorithm \ClockSynch\\
{\bf at} ``$\pulse$'' event \hfill\textit{/* received the internal pulse event */}\\
{\bf begin} \\
\mbox{$\,$\ }1. $Clock      := ET;$\\
\mbox{$\,$\ }2. Revoke possible other instances of \ClockSynch\ and\\
\tre            clear all data structures besides $ET$ and $Clock;$\\
\mbox{$\,$\ }3. Wait until $\sigma(1+ \rho)$ time units have elapsed since $\pulse;$ \\
\mbox{$\,$\ }4. $Next\_ET$  := \ByzConsensus($(ET + \Cycle)\bmod M,\;\sigma$)$;$\\
\mbox{$\,$\ }5. $Clock      := (Clock + Next\_ET-(ET+\Cycle)) \bmod M;$ \hfill\textit{/* posterior adjust. */}\\
\mbox{$\,$\ }6. $ET         := Next\_ET;$\\
{\bf end}
\normalsize
\end{minipage} }
\caption{The self-stabilizing Byzantine clock synchronization
 algorithm} \label{alg:pbss}
\end{figure}

The internal pulse event is delivered by the pulse synchronization
procedure. We assume the use of the pulse synchronization presented
in \cite{NEW-PULSE-TR}, though any pulse synchronization algorithm
that delivers synchronized pulses by solving the ``Self-stabilizing
Pulse Synchronization Problem'', in the presence of at most $f$
Byzantine nodes, where $n \geq 3f+1,$ such as the pulse procedure in
\cite{bio-pulse-synch}, can be executed in the background.

The pulse event aborts any possible on-going invocation of
\ClockSynch\ (and thus any on-going instant of \ByzConsensus) and
resets all buffers. The synchronization of the pulses ensures that
the \ClockSynch\ procedure is invoked within $\sigma$ real-time
units of its invocation at all other correct nodes.

Line 1 sets the local clock to the pre-agreed time associated with
the current pulse event. Line 3 intends to make sure that all
correct nodes invoke \ByzConsensus\ only after the pulse has been
invoked at all others, without remnants of past invocations, which
are revoked at Line 2. Past remnants may exist only during or
immediately following periods in which the system is not coherent.

In Line 4 \ByzConsensus\ intends to reach consensus on the next
value of $ET.$ One can use a synchronous consensus algorithm with
rounds of size $(\sigma+d)(1+2\rho)$ or asynchronous style consensus
in which a node waits to get $n-f$ messages of the previous round
before moving to the next round. We assume the use of a Byzantine
consensus procedure tolerating $f$ faults when $n \ge 3f+1.$ A
correct node joins \ByzConsensus\ only concomitant to an internal
pulse event, as instructed by the \ClockSynch. This contains the
possibility of faulty nodes to initiate consensus at arbitrary
times.

Line 5 is a posterior clock adjustment. It increments the clock
value with the difference between the agreed time associated with
the next pulse and the node's pre-consensus estimate for the time
associated with the next pulse (the value which it entered the
consensus with). This is equivalent to incrementing the value of
$ET$ that the node was supposed to hold at the pulse according to
the agreed $Next\_ET$ with the elapsed time from the pulse and until
the termination of \ByzConsensus. This intends to expedite the time
to reach synchronization of the clocks. In case that the
clock\_state before Line 5 was not a synchronized clock\_state then
a synchronized clock\_state is attained following termination of
\ByzConsensus\ at all correct nodes, rather than at the next pulse
event. Note that in the case that all correct nodes hold the same
$ET$ value at the pulse, then the posterior clock adjustment adds a
zero increment to the clock value.

Note that when the system is not yet coherent, following a chaotic
state, pulses may arrive to different nodes at arbitrary times, and
the $ET$ values and the clocks of different nodes may differ
arbitrarily. At that time not all correct nodes will join
\ByzConsensus\ and no consistent resultant value can be guaranteed.
Once the pulses synchronize (guaranteed by the pulse synchronization
procedure to happen within a single cycle) all correct nodes will
join the same instant of \ByzConsensus\ and will agree on the clock
value associated with the next pulse. From that time on, as long as
the system stays coherent the clock\_state remains a synchronized
clock\_state.

The use of Byzantine consensus tackles the clock wrap-around in a
trivial manner at all correct nodes.

Note that instead of simply setting the clock value to $ET$ we could
use some \emph{Clock-Adjustment} procedure (cf.~\cite{R3}), which
receives a parameter indicating the target value of the clock. The
procedure runs in the background, it speeds up or slows down the
clock rate to smoothly reach the adjusted value within a specified
period of time. This procedure should also handle the clock wrap
around.

\begin{theorem}\label{thm:pulse-cs}
\ClockSynch\ solves the ``Self-stabilizing Byzantine Clock
Synchronization Problem''.
\end{theorem}

\begin{proof}

\noindent{\bf Convergence:} Let the system be coherent but in an
arbitrary state $s,$ with the nodes holding arbitrary clock values.
Consider the first correct node that completed line $3$ of the
$\ClockSynch$ algorithm. Since the system is coherent, all correct
nodes invoked the preceding $\pulse$ within $\sigma$ of each other.
At the last pulse all remnants of previously invoked instances of
$\ByzConsensus$ were flushed by all the correct nodes. A correct
node does not initiate or join procedure \ByzConsensus\ before
waiting $\sigma(1+\rho)$ time units subsequent to the pulse, hence
not before all correct nodes have invoked a pulse and subsequently
flushed their buffers. Thus all correct nodes will eventually join
\ByzConsensus, thus \ByzConsensus\ will initiate and terminate
successfully.

At termination of the first instance of \ByzConsensus\ following the
synchronization of the pulses, all correct nodes agree on the clock
value to be associated with the next pulse invocation. Subsequently,
all correct nodes adjust their clocks, post factum, according to the
agreed $ET.$ Note that this posterior adjustment of the clocks does
not affect the time span until the invocation of the next pulse but
rather updates the clocks concomitantly to and in accordance with
the newly agreed $ET.$ This has an effect only if the correct nodes
joined \ByzConsensus\ with differing values. Hence if all correct
nodes join \ByzConsensus\ with the same $ET$ then the adjustment
equals zero. Since all correct pulses arrived within $\sigma$
real-time units of each other, after the posterior clock adjustment
of the last correct node, all correct clocks values are within
$$\gamma_{1}=\sigma(1+\rho)+(\sigma+\mbox{\slshape agreement\_duration})\cdot
2\rho$$ of each other. The $2\rho$ is the maximal drift rate between
any two correct clocks (whereas $\rho$ is their drift with respect
to real-time). Observe that $\gamma_{1} \le \gamma$ and therefore
the state of the system is a synchronized clock\_state. This
concludes the Convergence condition. \\ \qed

\noindent{\bf Closure:} Recall that system coherence is defined as a
continuous non-faulty behavior of the communication network and a
large enough fraction of the nodes for at least some minimal period
of time. The proof of the Closure condition assumes the correct
nodes have synchronized their $ET$ values, thus setting this minimal
time to be at least $\cyclemax + \mbox{\slshape
agreement\_duration}$ time, ensuring synchronization of the
variables.

Let the system be in a synchronized clock\_state and w.l.o.g. assume
all correct nodes hold synchronized and identical $ET$ values.
Observe that although the correct nodes have synchronized their $ET$
values this does not necessarily imply all correct nodes hold the
same $ET$ value at every point in time. At a brief time subsequent
to the termination of \ByzConsensus, only a part of the correct
nodes may have set the $ET$ to the new agreed value while the rest
of the correct nodes currently holding the old $ET$ value will set
$ET$ to the new value in a brief time. We first prove the first
Closure condition (\textit{precision}). In this case, each correct
node adjusts its clock immediately subsequent to the pulse, but the
posterior clock adjustment has no effect since the consensus value
equals the value it joined \ByzConsensus\ with. To simplify the
discussion assume for now that no wrap around of any correct clock
takes place during the time that the pulse arrives at the first
correct node and until the pulse is invoked at the last correct
node. Immediately after the pulse is invoked at the last correct
node and its subsequent clock adjustment, all correct clocks are
within $\gamma_{0}=\sigma(1+\rho)$ of each other.

From that point on, clocks of correct nodes drift apart at a rate of
$2\rho$ of each other. As long as no wrap around of the clocks takes
place and no pulse arrives at any correct node, the clocks are at
most $\gamma_0+\Delta T\cdot 2\rho$ apart, where $\Delta T$ is the
real-time elapsed since the invocation of the pulse at the first
correct node. To estimate the maximal clock difference, $\gamma,$ at
any time, we will consider the following complementary cases:
\begin{itemize}
  \item[P1)~] Prior to the next pulse event at the first correct node.
  \item[P2)~] When a pulse arrives at some correct node.
  \item[P3)~] Immediately after the last node invokes its next pulse event.
\end{itemize}

Note that in this case we do not need to consider the posterior
adjustment of the clocks at Line 5.

Case P1 cannot last more than $\Delta T=\cyclemax,$ since by the end
of that time interval all correct nodes will have invoked the pulse,
reducing to case P2 or P3. The discussion above implies
$\gamma=\gamma_0+\cyclemax\cdot 2\rho.$

Case P3 implies that clock readings are at most $\gamma_0$ apart,
since all nodes invoke the pulses within $\sigma.$

To analyze case P2 consider that the next pulse event has been
invoked at some node, $p.$ The following situations may take
place:
\begin{itemize}
  \item[P2a)] Following its clock adjustment, the clock of $p$ holds
  the maximal clock value among all correct clocks at that moment.
  \item[P2b)] Following its clock adjustment, the clock of $p$ holds
  the minimal clock value among all correct clocks at that moment.
  \item[P2c)] Neither of the above.
\end{itemize}

In case P2a, since $p$ holds the maximal clock value, we claim that
no other clock reading can read less than
$ET_{\mbox{\scriptsize\slshape lastpulse}}+\cyclemin\cdot(1-\rho).$
Assume by contradiction the existence of a correct node $q$ whose
clock reading is less than this value. Further assume that node $q$
received the same set of messages from the same sources and at the
same time as node $p.$ These events caused node $p$ to invoke its
pulse and would necessarily cause node $q$ to also invoke a pulse.
The elapsed time on the clock of node $q$ between the current pulse
and the previous is thus less than $\cyclemin\cdot(1-\rho)$ which is
less than $\cyclemin$ real-time after its previous pulse. A
contradiction to the definition of $\cyclemin.$ Node $p$ just
adjusted its clock which thus reads
$ET=ET_{\mbox{\scriptsize\slshape lastpulse}}+\Cycle.$ Due to the
clock skew the clock difference may increase an additional
$2\rho\sigma$ until the node invokes its pulse and the case reduces
to P3. The discussion above implies
$\gamma=(ET_{\mbox{\scriptsize\slshape
lastpulse}}+\Cycle)-(ET_{\mbox{\scriptsize\slshape
lastpulse}}+\cyclemin\cdot
(1-\rho))+2\rho\sigma=\Cycle-\cyclemin\cdot(1-\rho)+2\rho\sigma.$

In case P2b, the clock readings of all other nodes that have invoked
a pulse can not be more than $\gamma_0$ apart (case P3). The clock
reading of any node that has not invoked a pulse yet should be less
than $\cyclemax$ following similar reasoning as in case P2a. Node
$p$ just adjusted its clock which thus reads
$ET=ET_{\mbox{\scriptsize\slshape lastpulse}}+\Cycle.$ Due to the
clock skew the clock difference may increase an additional
$2\rho\sigma$ until the node invokes its pulse and the case reduces
to P3. The discussion above implies
$\gamma=(ET_{\mbox{\scriptsize\slshape lastpulse}}+\cyclemax\cdot
(1+\rho))-(ET_{\mbox{\scriptsize\slshape
lastpulse}}+\Cycle)+2\rho\sigma=\cyclemax\cdot(1+\rho)-\Cycle+2\rho\sigma.$

For case P2c, if the nodes holding the minimal clock reading and
maximal clock reading already invoked pulses, then the clock
difference reduces to case P3.

If neither of the nodes holding the minimal and maximal clock
values have not invoked their pulses yet, then the clock
difference reduces to case P1.

Otherwise, if either the node holding the minimal or the maximal
clock value already invoked its pulse then one of the bounds of
P2a or P2b hold until the other node invokes its pulse.

We now consider the case that a clock wrap around takes place at
some $\Delta T$ real-time after the last pulse is invoked in the
synchronized cycle. From the discussion earlier we learn that at the
moment prior to the first correct clock wraps around, the correct
clocks are at most $\gamma$ apart. Therefore, all correct clocks
will wrap around within at most another $\gamma$ time. During the
intermediate time, any two correct clocks, $i,j,$ for which one has
wrapped around and the other not, satisfy $|clock_i(t)-clock_j(t)|
\ge M -\gamma.$ Thus we proved that the maximal clock difference
will remain less than $\gamma$ or greater than $M -\gamma,$ which
completes the first Closure condition.

Henceforth, the bound on the clock differences of correct nodes will
equal the maximal of the three values calculated above. Formally
this yields $\gamma=$
$max[\cyclemax\cdot(1+\rho)-\Cycle+2\rho\sigma,\;
\Cycle-\cyclemin\cdot(1-\rho)+2\rho\sigma,\;
\sigma(1+\rho)+\cyclemax\cdot 2\rho].$ The explicit value is
dependent on the relationship between $\cyclemax,$ $\cyclemin$ and
$\Cycle,$ which is determined by the pulse synchronization procedure
(\cite{NEW-PULSE-TR}). The explicit value of $\gamma$ is presented
in Section~\ref{sec:analysis}. This concludes the first Closure
condition.

For the second Closure condition, note that $\Psi_i,$ as defined in
Section~\ref{sec:model}, represents the actual deviation of an
individual correct clock ($p_i$) from the real-time interval during
which it progresses. This is equivalent to the maximal actual
difference between the clock value and real-time during a real-time
interval in which real-time and the clock value were equal at the
beginning of the interval. The \textit{accuracy} of the clocks is
the bound on the actual deviation of correct clocks from any finite
real-time interval or rate of deviation from the progression of
real-time. Thus it suffices to show that correct clocks progress
with an accuracy that is a linear function of every finite real-time
interval to satisfy the second Closure condition.

The clock progression has an inherent deviation from any real-time
interval due to the physical clock skew. In addition, the clocks are
repeatedly adjusted at every pulse in order to tighten the
precision, which can further deviate the clocks progression from the
progression of the real-time during the interval. In
\cite{NEW-PULSE-TR} it is shown that the pulses progress with a
linear envelope of any real time interval. The accuracy in a cycle
equals the bound on the clock adjustment
$|t_{\mbox{\scriptsize\slshape pulse}}-ET_{\mbox{\scriptsize\slshape
pulse}}|,$ where $t_{\mbox{\scriptsize\slshape pulse}}$ is the clock
value at the pulse at the moment prior to the adjustment of the
clock to $ET_{\mbox{\scriptsize\slshape pulse}}.$ Under perfect
conditions, i.e. no clock skew and zero clock adjustment
$t_{\mbox{\scriptsize\slshape pulse}}=ET_{\mbox{\scriptsize\slshape
pulse}}.$ This would further equal real-time should the clocks have
initiated with real-time values. Thus it suffices to show that the
adjustment to the clocks at every pulse is a linear function of the
length of the cycle. The upper and lower bounds on the value
$t_{\mbox{\scriptsize\slshape pulse}}$ is determined by the bound on
the effective cycle length and accounts for the clock skew and the
accuracy of the pulses (bound on the deviation of the pulses from
perfect regularity). Let $\cyclemin$ and $\cyclemax$ denote the
lower bound and upper bound respectively on the cycle length in
real-time units. Hence,
$$ET_{\mbox{\scriptsize\slshape prev-pulse}} + \cyclemin\cdot(1-\rho) \le t_{\mbox{\scriptsize\slshape pulse}} \le
ET_{\mbox{\scriptsize\slshape prev-pulse}} +
\cyclemax\cdot(1+\rho)\;.$$ The adjustment
to the correct clocks, $ADJ,$ is thus bounded by\\

\noindent$ET_{\mbox{\scriptsize\slshape pulse}} -
[ET_{\mbox{\scriptsize\slshape prev-pulse}} +
\cyclemax\cdot(1+\rho)] \le 0
 \le ADJ \le 0$ \\ {\flushright $ \le ET_{\mbox{\scriptsize\slshape pulse}} - [ET_{\mbox{\scriptsize\slshape prev-pulse}} +
\cyclemin\cdot(1-\rho)]\;,$ \\} {\flushleft which translates to\\}
\noindent{\flushleft$ET_{\mbox{\scriptsize\slshape prev-pulse}}+
\Cycle -[ET_{\mbox{\scriptsize\slshape prev-pulse}}+
\cyclemax\cdot(1+\rho)] \le ADJ \le$\\} {\flushright
$ET_{\mbox{\scriptsize\slshape prev-pulse}}+ \Cycle
-[ET_{\mbox{\scriptsize\slshape
prev-pulse}}+\cyclemin\cdot(1-\rho)]\;,$\\} {\flushleft which
translates to}
\\
$$\Cycle - \cyclemax\cdot(1+\rho) \le ADJ \le
\Cycle - \cyclemin\cdot(1-\rho)\;.$$

As can be seen, the bound on the adjustment to the clock is linear
in the effective cycle length. The bounds on the effective cycle
length are guaranteed by the pulse synchronization procedure to be
linear in the default cycle length. Thus the accuracy of the clocks
are within a linear envelope of any real-time interval. The actual
values of $\cyclemin$ and $\cyclemax$ are determined by the specific
pulse synchronization procedure used. This concludes the Closure
condition. \\ \qed

Thus the algorithm is self-stabilizing and performs correctly with
$f$ Byzantine nodes for $n \geq 3f+1.$ \qed

\end{proof}

\vspace{-2mm} \noindent
\subsection{A Clock Synchronization
Algorithm without Consensus} \vspace{-2mm} We suggest a simple
additional Byzantine self-stabilizing clock synchronization
algorithm using pulse synchronization as a building block that does
not use consensus.

Our second algorithm resets the clock at every pulse\footnote{This
approach has been suggested by Shlomi Dolev as well.}. This approach
has the advantage that the nodes never need to exchange and
synchronize their clock values and thus do not need to use
consensus. This version is useful for example when $M,$ the
upper-bound on the clock value, is relatively small. The algorithm
has the disadvantage that for a large value of $M,$ a large $\Cycle$
value is required. This enhances the effect of the clock skew, thus
negatively affecting the precision and the accuracy at the end of
the cycle. Note
that the precision and accuracy of \CYCLECS equals that of \ClockSynch.\\

\vspace{-4mm}\noindent
\begin{figure}[!h]
\center \fbox{\begin{minipage}{4.5in} \footnotesize
\setlength{\baselineskip}{3.5mm}

Algorithm \CYCLECS\\
{\bf at} ``$\pulse$'' event \hfill\textit{/* received the internal pulse event */}\\
{\bf begin} \\
\tb     $Clock := 0;$\\
{\bf end}
\normalsize
\end{minipage} }
\caption{Additional CS algorithm in which the clock wraps-around
every cycle } \label{alg:cycle-cs}
\end{figure}

\vspace{-2mm} \noindent
\subsection{A Clock Synchronization
Algorithm using an Approximate Agreement Approach} \vspace{-2mm} We
suggest an additional self-stabilizing Byzantine clock
synchronization algorithm using pulse synchronization as a building
block, denoted \ApproxClocks.

The algorithm uses an approximate agreement approach in order to get
continuous clocks with high precision and accuracy on expense of the
message complexities and early-stopping property. The precision and
the accuracy are $2\sigma+O(\rho)$ and thus improve on those of
\ClockSynch.

\begin{figure}[!h]
\center \fbox{\begin{minipage}{4.92in} \footnotesize
\setlength{\baselineskip}{3.5mm}

Algorithm \ApproxClocks\\
{\bf at} ``$\pulse$'' event \hfill\textit{/* received the internal pulse event */}\\
{\bf begin} \\
\mbox{$\,$\ }1. $Clock$-$at$-$pulse:=Clock;$\\
\mbox{$\,$\ }2. Revoke possible other instances of $\ApproxClocks$ and\\
\tre            clear all data structures besides $Clock$-$at$-$pulse;$\\
\mbox{$\,$\ }3. Wait until $\sigma(1+ \rho)$ time units have elapsed since $\pulse;$ \\
\mbox{$\,$\ }4. $Clock_{Consensus}$  := \ApproxByzAgreement(\mbox{$Clock$-$at$-$pulse$})$;$\\
\mbox{$\,$\ }5. $Clock      := (Clock_{Consensus} + \mbox{elapsed-time-since-pulse}) \bmod M;$ \\
{\bf end}
\normalsize
\end{minipage} }
\caption{Self-stabilizing Byzantine Approximate Clock
Synchronization algorithm} \label{alg:approx-cs}
\end{figure}

In Line 4 of \ApproxClocks\ the nodes invoke approximate-like
agreement on their local clock value at the time of the last pulse,
denoted Clock-at-pulse. In case that the system state was a
synchronized clock\_state then the resultant value
$Clock_{Consensus}$ is guaranteed by the \ApproxByzAgreement\ to be
in the range of the initial clock values of the correct nodes. If
the clocks were not synchronized then the resultant agreed value
 may be in any range. In Line 5 every correct node
sets its clock to equal the agreed clock value associated with the
last pulse, $Clock_{Consensus},$ incremented with the time that has
elapsed on its local timer since the pulse.

\begin{figure}[!h]
\center \fbox{\begin{minipage}{5.1in} \footnotesize
\setlength{\baselineskip}{3.5mm}

Algorithm \ApproxByzAgreement(value)\\
{\bf begin} \\
\mbox{$\,$\ }1. Invoke \ByzAgreement() on $value;$\\
\mbox{$\,$\ }2. After termination of all \ByzAgreement\ instances (substitute missing values with $0$) Do:\\
\mbox{$\,$\ }3. Find largest set of values within $\gamma$+$\sigma$ of each other (if several, choose set harboring smallest value $\ge 0$)$;$\\
\mbox{$\,$\ }4. Find median of the set, identify its antipode $:= (\mbox{median}+\lfloor M/2\rfloor) \bmod M;$\\
\mbox{$\,$\ }5. Discard the $f$ immediate values from each side of the antipode$;$ \\
\mbox{$\,$\ }6. Return the median of the remaining values$;$\\
{\bf end}
\normalsize
\end{minipage} }
\caption{Self-stabilizing Byzantine Approximate Agreement}
\label{alg:approx-agree}
\end{figure}

In order to be self-contained we bring the definition of
\emph{Approximate Agreement}, defined in \cite{Rapprox}.

Formally, the goal of $\epsilon$-Approximate Agreement is to reach
the following: let there be $n$ processes $p_1, . . . , p_n,$ each
starts with an initial value $v_i \in \mathbb{R}$ and may decide on
a value $d_i \in \mathbb{R}.$

\begin{enumerate}

\item \textbf{(Approximate Agreement)} If $p_i$ and $p_j$ are
correct and have decided then $|d_i-d_j| \le \epsilon.$\\

\item \textbf{(Validity)} If $p_i$ is correct and has decided
then there exists two correct nodes $p_j,p_k$ such that $v_j \le d_i
\le v_k,$ (the decision value of every correct node is in the range
of the initial values of the correct nodes).\\

\item \textbf{(Termination)} All correct nodes eventually decide.\\

\end{enumerate}

The approximate agreement protocol in \cite{Rapprox} cannot be used
as-is in the self-stabilization model as the notions of ``highest''
value and ``lowest'' value are not defined when nodes can initialize
with values reaching their bounds, $M.$ Faulty nodes can in this
case cause different correct nodes to view the extremes of the
values as complete opposites. To overcome the lack of total order
relation introduced by the self-stabilization model,
\ApproxByzAgreement\ thus combines the approximate agreement
algorithm of \cite{Rapprox} with Byzantine agreement as follows: run
separate Byzantine agreements in parallel on every node's value in
order to agree on the value of each node. Thus all correct nodes
will hold identical multisets and henceforth the heuristics of
\cite{Rapprox} will be executed on exactly the same values at all
correct nodes. The \ApproxByzAgreement\ procedure satisfies the
conditions for classic approximate agreement, while being
self-stabilizing.

The \ByzAgreement\ procedure used is the Byzantine agreement of
\cite{FastAgree87}, though using our \Broadcast primitive presented
in Section~\ref{sec:bcast-primitive} in order to overcome the lack
of any common reference to clock time among the correct nodes.

In Line 1 of \ApproxByzAgreement, every node invokes Byzantine
agreement on its value, within $\sigma$ real-time of each other.
Every instance of $\ApproxByzAgreement$ must terminate within some
bounded time, thus all correct nodes can calculate a time when all
the agreement instances have terminated at all correct nodes. In
Line 3, after all the agreement instances have terminated and
missing values are substituted with a $0,$ a set of supposedly
synchronized values is searched for. Note that if not all instances
of $\ApproxByzAgreement$ have terminated within the pre-calculated
time-bound then the system must have been in a non-coherent state.
Synchronized clock values can be up-to $\gamma+\sigma$ apart in the
values agreed subsequent to Line2, due to the pulse uncertainty. In
Line 4 the median of the set is identified, and will serve as an
anchor for determining the order relation among the different
values. In Line 5, the antipode (in the range $1..M$) of the median
is identified; the $f$ first values on each side of this antipode
are then discarded. If the system is in a synchronized clock\_state
then all values that are outside of the values in the set identified
earlier are discarded. Thus the median of the remaining values,
returned in Line 6, is in the range of the initial values of the
correct nodes.

\begin{lemma}The \ApproxByzAgreement\ procedure satisfies all the
conditions for $\epsilon$-Approximate Agreement, for $\epsilon=0,$
when the system is in a synchronized clock\_state\footnote{The
notion ``in the range of'' remains undefined if the system is not in
a synchronized clock\_state. Thus the validity condition remains
undefined for this case.}.
\end{lemma}

\begin{proof} Note the validity of $\ByzAgreement$
guarantees that the value decided by all correct nodes for node $i$
is $i$'s actual input value.

\begin{enumerate}

\item \textbf{Approximate\_Agreement}: All correct nodes hold the
same multiset of values following all terminations of the instances
of \ByzAgreement, thus they all find the same set in Line 3 and
hence do the exact same operations in lines 3-5, and thus return the
same value in Line 6.

\item \textbf{Validity}: Let the system be in a synchronized clock\_state.
Thus the agreed clock values for all correct nodes subsequent to
executing Line 2 are at most $\gamma+\sigma$ apart. Hence, the
largest set found in Line 3 includes at least $n-f$ values. We now
seek to prove that the decision value is in the range of the initial
values of the correct nodes. Since $f<n/3$ it follows that all
values that are not in the range (at most $f$) of this set are
discarded in Line 5. Thus all remaining values must be in the range
of the initial values of the correct nodes. In particular, the
median of the remaining values is in the range of the initial
values. This completes the proof of the validity condition.

\item \textbf{Termination}: Follows from the termination of \ByzAgreement.

\end{enumerate}
\qed
\end{proof}

The precision $\gamma,$ is the bound on the clock differences of all
correct nodes at any time.

\begin{lemma} The precision of \ApproxByzAgreement\ is $2\sigma + O(\rho).$
\end{lemma}

\begin{proof} At the moment after all correct nodes have executed Line
5 in $\ApproxClocks$ their clocks differ by at most
$\sigma+O(\rho),$ thus the clock differences are at most
$\sigma+O(\rho)$ also at the forthcoming pulse invocation. The
precision $\gamma,$ is maximized at the moment that a correct node
has set its clock subsequent to its execution of Line 5 in
\ApproxByzAgreement, while some other node has yet to execute this
line. Following the validity condition, the agreed clock value
$Clock_{Consensus},$ is within the initial clock values that was
held by the correct nodes at their last pulse. As the system is in a
synchronized clock\_state thus these initial values were within
$2\sigma+O(\rho)$ of each other. Thus the node that has just
adjusted its clock, set it to a value that is within
$2\sigma+O(\rho)$ of its clock at the moment before the adjustment.
In particular this adjusted clock value is also within
$2\sigma+O(\rho)$ of the clock value of any other correct node. This
observation yields a precision of $\gamma=2\sigma+O(\rho).$

\qed
\end{proof}

The accuracy equals the maximal clock adjustment which for the same
arguments as above yields an accuracy of $2\sigma+O(\rho).$

A self-stabilizing Byzantine approximate agreement algorithm that
knows how to handle bounded, wrapping values and thus does not need
to reach exact agreement on every node's value, will supposedly
yield a clock synchronization algorithm with time and message
complexity comparable to $\ClockSynch$ with precision and accuracy
of \ApproxClocks. To the best of our knowledge no such approximate
agreement algorithm exists.

\section{Analysis and Comparison to other Clock Synchronization Algorithms }
\label{sec:analysis} \vspace{-2mm}

Our clock synchronization algorithm \ClockSynch\ requires reaching
consensus in every cycle. This implies that the cycle should be long
enough to allow for the consensus procedure to terminate at all
correct nodes. This implies having $\cyclemin\ge 2\sigma+3(2f+4)d,$
assuming that the \ByzConsensus\ procedure takes $(f+2)$ rounds of
$3d$ each. The algorithm has the advantage that it uses the full
time to reach consensus only following a catastrophic state in which
correct nodes hold differing $ET$ values. Once in a synchronized
clock\_state, all correct nodes participate in the consensus with
the same initial consensus value which thus terminates within 2
communication rounds only, due to its early-stopping property.
Hence, during steady state, in which the system is in a legal state,
the time and message complexity overhead of \ClockSynch\ is minimal.

For simplicity we also assume $M$ to be large enough so that it
takes at least a cycle for the clocks to wrap around.

Note that $\Psi_i,$ defined in Section~\ref{sec:model}, represents
the actual deviation of an individual correct clock, $p_i,$ from a
given real-time interval. The \emph{accuracy} of the clocks is the
bound on this deviation of correct clocks from any real-time
interval. The clocks are repeatedly adjusted in order to minimize
the accuracy. Following a synchronization of the clock values, that
is targeted to occur once every $\Cycle$ time units, correct clocks
can be adjusted by at most $ADJ,$ where following
Theorem~\ref{thm:pulse-cs},
$$\Cycle - \cyclemax\cdot(1+\rho) \le ADJ \le
\Cycle - \cyclemin\cdot(1-\rho)\;,$$ which, following $\cyclemin$
and $\cyclemax$ determined by the pulse synchronization procedure of
\cite{NEW-PULSE-TR} to equal $\Cycle-11d$ and $\Cycle+9d$
respectively, translates to

$$ -9d(1+\rho) -\rho\cdot\Cycle
\le ADJ \le 11d(1-\rho)+ \rho\cdot\Cycle\;.$$

\ignore{$$ -3d(1+\rho) -2\rho\frac{\Cycle}{1 - \rho} \le ADJ \le
2d(1-\rho)+ 2\rho\frac{\Cycle}{1 + \rho}\;.$$}

The accuracy is thus $11d+O(\rho)$ real-time units. Should the
initial clock values reflect real-time then this determines the
accuracy of the clocks with respect to real-time (and not only with
respect to real-time progression rate), as long as the system is
coherent and clocks do not wrap around.

Recall that the precision $\gamma,$ is the bound on the difference
between correct clock values at any time. This bound is largely
determined by the maximal clock value difference at the time in
which a correct node has just set its clock and some other correct
node is about to do it in a short time. It is guaranteed by
Theorem~\ref{thm:pulse-cs} and the pulse synchronization tightness
$\sigma=3d$ of \cite{NEW-PULSE-TR}, to be:

\begin{eqnarray*}
\gamma &=& \max[\cyclemax\cdot(1+\rho)-\Cycle+2\rho\sigma,\\
&&\mbox{\ \ \ \ \ \ \ \ }\Cycle-\cyclemin\cdot(1-\rho)+2\rho\sigma,\;\sigma(1+\rho)+\cyclemax\cdot 2\rho]\\
&=&\max[9d(1+\rho)+ \rho\cdot\Cycle+2\rho\sigma,\;
11d(1-\rho)+ \rho\cdot\Cycle+2\rho\sigma,\\
&&\mbox{\ \ \ \ \ \ \ \ }3d(1+\rho)+(\Cycle + 9d)\cdot 2\rho]\\
&=&11d(1-\rho)+ \rho\cdot\Cycle+2\rho\sigma=11d+O(\rho)\;.
\end{eqnarray*}\\

The bound on the difference between correct clock values immediately
after all correct nodes have synchronized their clock value (at Line
1 or Line 5) is $\sigma.$

\begin{table}[t!]
\label{Table1}
 \noindent{\footnotesize
\begin{tabular}{|l|c|c|c|c|c|}
  \hline
   Algorithm & Self-        & Precision  & Accuracy       & Convergence  & Messages\\
             & stabilizing  & $\gamma$   &                & Time         & \\
             & /Byzantine   &            &                &              &\\
  \hline
\ClockSynch&SS+BYZ& $11d + O(\rho)$  &
$11d + O(\rho)$ & $\cyclemax~+~$ & $O(nf^{2})$\\
& & & &~~ $3(2f+5)d$ & \\
\CYCLECS& SS+BYZ &  $11d + O(\rho)$ &
$11d + O(\rho)$ & $\cyclemax$ & $O(n^{2})$\\
\ApproxClocks& SS+BYZ &  $3d + O(\rho)$ &
$3d + O(\rho)$ & $\cyclemax$ & $O(nf)^{2}$\\
\mbox{{\scriptsize DHSS~\cite{R3}}}& BYZ & $d+O(\rho)$  &
$(f+1)d+ O(\rho)$& $2(f+1)d$ & $O(n^{2})$\\
\mbox{{\scriptsize LL-APPROX~\cite{WelchLynch88}}}& BYZ &
$5\epsilon + O(\rho)$ &  $\epsilon + O(\rho)$ & $d+O(\epsilon)$ & $O(n^{2})$\\
\mbox{{\scriptsize DW-SYNCH~\cite{DolWelSSBYZCS04}}*}&SS+BYZ& 0
& 0  & $M2^{2(n-f)}$ & $n^{2}M2^{2(n-f)}$\\
\mbox{{\scriptsize DW-BYZ-SS~\cite{DolWelSSBYZCS04}}}&SS+BYZ&
$4(n-f)\epsilon+O(\rho)$ & $(n-f)\epsilon+O(\rho)$ & $O({n)^{O(n)}}$
& $O({n)^{O(n)}}$\\
\mbox{{\scriptsize PT-SYNC~\cite{PT97}}*}& SS & 0 &
0 & $4n^{2}$ & $O(n^{2})$\\
  \hline
\end{tabular}
} \caption{Comparison of clock synchronization algorithms
($\epsilon$ is the uncertainty of the message delay). The
convergence time is in pulses for the algorithms utilizing a global
pulse system and in rounds for the other semi-synchronous protocols.
PT-SYNC assumes the use of shared memory and thus the ``message
complexity'' is of the ``equivalent messages''. The '*' denotes the
use of a global pulse or global clock tick system.}
\end{table}

The only self-stabilizing Byzantine clock synchronization
algorithms, to the best of our knowledge, are published in
~\cite{R22,DolWelSSBYZCS04}. Two randomized self-stabilizing
Byzantine clock synchronization algorithms are presented, designed
for fully connected communication graphs, use message passing which
allow faulty nodes to send differing values to different nodes,
allow transient and permanent faults during convergence and require
at least $3f+1$ processors. The clocks wrap around, where $M$ is the
upper bound on the clock values held by individual processors. The
first algorithm assumes a common global pulse system and
synchronizes in expected $M \cdot 2^{2(n-f)}$ global pulses. The
second algorithm in~\cite{DolWelSSBYZCS04} does not use a global
pulse system and is thus partially synchronous similar to our model.
The convergence time of the latter algorithm is in expected
$O((n-f){n^{6(n-f)}})$ time. Both algorithms thus have drastically
higher convergence times than ours.

In Table 1 we compare the parameters of our protocols to previous
classic Byzantine clock synchronization algorithms, to non-Byzantine
self-stabilizing clock synchronization algorithms and to the prior
Byzantine self-stabilizing clock synchronization algorithms. It
shows that our algorithm achieves precision, accuracy, message
complexity and convergence time similar to non-stabilizing
algorithms, while being self-stabilizing.

The message complexity of \ClockSynch\ is solely based on the
underlying Pulse and Consensus procedures. Its inherent convergence
time is $\cyclemax.$ The $O(nf^{2})$ message complexity as well as
the $+3(2f+5)d$ additive in the convergence time come from
\ByzConsensus, the specific Byzantine consensus procedure we use.
The pulse synchronization procedure we use from \cite{NEW-PULSE-TR}
has a message complexity of $O(n^{2})$ and $6\cdot \cycle$
convergence time. Note that \ByzConsensus\ has two early-stopping
features: It stops in a number of rounds dependent on the actual
number of faults and if nodes initiate with the same values (same
$ET$ values) then it stops within 2 rounds.

Note that some of the algorithms cited in Table 1 refer to
$\epsilon,$ the uncertainty in message delivery, rather than $d,$
the end-to-end communication network delay.

The DW-SYNCH and PT-SYNCH algorithms cited in Table 1 make use of
global clock ticks (common physical timer). Note that this does not
make the clock synchronization problem trivial as such clock ticks
can not be used to invoke agreement procedures and the nodes still
need to agree on the clock values. The benefit of utilizing a global
pulse systems is in the optimal precision and accuracy acquired (see
\cite{DolWelSSBYZCS04}).

\ignore{ Note that if instead of using the pulse synchronization
procedure of \cite{NEW-PULSE-TR}, one uses the pulse synchronization
of~\cite{bio-pulse-synch} then the precision can improve, but the
convergence time would drastically increase.}


\ignore{-2mm}

\appendix
\section{Appendix - The Consensus and Broadcast Primitives}\label{sec:apdx}

\subsection{The \ByzConsensus\ Procedure} \label{sec:byzssconsensus}
\vspace{-2mm}The \ByzConsensus\ procedure can implement many of the
classical Byzantine consensus algorithms. It assumes that timers of
correct nodes are always within $\bsigma$ of each other. More
specifically, we assume that nodes have timers that reset
periodically, say at intervals $\le \cycle^\prime.$ Let $T_i(t)$ be
the reading of the timer at node $p_i$ at real time $t.$ We thus
assume that there exists a bound such that for every time $t,$ when
the system is coherent,
$$\forall i,j\;\;\mbox{if}\;\bsigma< T_i(t),T_j(t)< \cycle^\prime-\bsigma\;
\mbox{then}\;|T_i(t)-T_j(t)|< \bsigma\;.$$ The bound $\bar\sigma$
includes all drift factors that may occur among the timers of
correct nodes during that period. When the timers are reset to zero
it might be that, for a short period of time, the timers may be
further apart. The pulse synchronization algorithm in
\cite{NEW-PULSE-TR} satisfies the above assumptions and implies
$\bsigma\ge d.$

The self-stabilization requirement and the deviation that may arise
from any synchronization assumption imply that any consensus
protocol must be carefully specified. The consensus algorithm will
function properly if it is invoked when the timers of correct nodes
are within $\bsigma$ of each other. The subtle point is to make sure
that an arbitrary initialization of the procedure cannot cause the
nodes to block or deadlock. Below we show how to update the early
stopping Byzantine Agreement algorithm of Toueg, Perry and
Srikanth~\cite{FastAgree87} to become self-stabilization and to make
it into a general consensus (vs. agreement) procedure.

The procedure does not assume any reference to real-time and no
complete synchronization of the rounds, as is assumed
in~\cite{FastAgree87}. Rather it resets the local timers of correct
nodes at each pulse which thus makes the timers within bounds of
each other. The node invokes the procedure with the value to agree
on and the local timer value. In the procedure nodes also consider
all messages accumulated in their buffers that were accepted prior
to the invocation, if they are relevant.

We use the following notations in the description of the consensus
procedure: 
\begin{itemize}
\vspace{-2mm}

\item Let $\bar d$ be the duration of time equal to
$(\bsigma+ d)\cdot(1+\rho)$ time units on a correct node's timer.
Intuitively, $\bar d$ can be assumed to be a duration of a ``phase''
on a correct node's timer.

\item The \emph{\Broadcast} primitive is the primitive defined in
Section~\ref{sec:bcast-primitive} and is an adaptation of the one
described in~\cite{FastAgree87}. Note that an \emph{accept} is
issued within the \Broadcast primitive.

\end{itemize}

The main differences from the original protocol
of~\cite{FastAgree87} are:

\begin{itemize}

\vspace{-2mm} \item Instead of the General in the original protocol
we use a virtual (faulty) ``General'' notion of a virtual node whose
value is the assumed value of all correct nodes at a correct
execution. It is the value with which the individual nodes invoke
the procedure. Thus, every correct node does a \ConsBCast of its
initial $Val$ in contrast to the original protocol in which only the
General does this. If all correct nodes initiate with the same value
and at the same timer time this will be the agreed value.

\item The \ConsBCast primitive has been modified by omitting
the code dealing with the $init$ messages. All correct nodes send an
echo of their initial values as though they previously received the
$init$ message from the virtual General.

\item It is assumed that the \Broadcast and \ConsBCast primitives
are implicitly initiated when a corresponding message arrives.
\end{itemize}

\ByzConsensus\ is presented in a somewhat different style. Each step
has a condition attached to it, if the condition holds and the timer
value assumption holds, then the step is to be executed. Notice that
only the step needs to take place at a specific timer value.

\begin{figure}[!h]
\center \fbox{\begin{minipage}{4.6in} \footnotesize
\setlength{\baselineskip}{3.5mm}

\noindent Procedure {\bf \ByzConsensus}($Val,T$)
\mbox{\ }\hfill\textit{ /* invoked at $p$ with timer $T$ */}\\

\noindent$broadcasters:=\emptyset$; $value=\perp$;\\

\noindent Do {\bf \ConsBCast}$(General,Val,T,1);$\\

{\bf by time }$(T+2\bar d):$ \\
\tb {\bf if} \textbf{accepted} $(General,v,T,1)$
\textbf{then}\\\tre $value:=v;$\\

{\bf by time }$(T+(2f+4)\bar d):$ \\
\tb{\bf if} $value\ne \perp$ \textbf{then}\\
\tre \textbf{\Broadcast}
$(p,value,T,\lfloor\frac{T_i-T}{2\bar d}\rfloor+1);$\\
\tre \textbf{stop} and \textbf{return} $value.$ \\

{\bf at time }$(T+2r\bar d):$ \\
\tb{\bf if} ($|broadcasters| < r-1$)   \textbf{then}\\
\tre\textbf{stop} and \textbf{return} $value.$ \\

\bf by time $(T+2r\bar d):$ \\
\tb \textbf{if}  \textbf{accepted} $(General,v^\prime,T,1)$ and
$r-1$ distinct messages
$(q_i,v^\prime,T,i)$ \\
\tre\tre where $\forall i,j\;\;2\le i \le r,$ and $q_i\neq q_j$ \textbf{then}\\
\tre  $value:=v^\prime;$
\normalsize
\end{minipage} }
\caption{The \ByzConsensus\ procedure} \label{alg:Byz-alg}
\end{figure}

\newpage
The \ByzConsensus\ procedure satisfies the following typical
properties:

\begin{enumerate}

\item[ {\bf Termination:}] The protocol terminates in a finite time;

\item[ {\bf Agreement:}] The protocol returns the same
value at all correct nodes;

\item[ {\bf Validity:}] If all correct nodes invoke the protocol
with the same value and time, then the protocol returns that value;
\end{enumerate}

It also satisfies the following {\bf early stopping} properties:

\begin{enumerate}
\item[\bf ES-1]  If all correct nodes invoke the protocol with the same
consensus value and with the same timer value, then they all stop
within two ``rounds'' of information exchange among correct nodes.

\item[\bf ES-2] If the actual number of faults is $f^\prime \le f$
then the algorithm terminates by $\min[T+(2f^\prime+6)\bar
d,\;T+(2f+4)\bar d]$ on the timer of each correct node.
\end{enumerate}

Notice that [ES-1] takes in practice significantly less time than
the specified upper bound on the message delivery time.
\begin{figure}[!h]
\center \fbox{\begin{minipage}{4.5in} \footnotesize
\setlength{\baselineskip}{3.5mm}

\noindent Procedure {\bf \ConsBCast}$(General,v,\tau,1)$\\
\mbox{\ }\hfill\textit{/* invoking a broadcast simulating the General */}\\
\mbox{\ }\hfill\textit{/* nodes send specific message with the same $\tau$ only once */}\\
\mbox{\ }\hfill\textit{/* multiple  messages
 sent by an  individual node are ignored*/}\\

send $(echo,General,v,\tau,1)$ to all;\\

{\bf by time }$(\tau+\bar d):$ \\
\tb \textbf{if} received
$(echo,General,v,\tau,1)$ from $\ge n-2f$ distinct nodes \textbf{then}\\
\tre  $broadcasters:=broadcasters\bigcup\{General\}$ ;\\
\tb \textbf{if} received $(echo,General,v,\tau,1)$ from $\ge n-f$
distinct nodes $q$
\textbf{then}\\
\tre send $(echo^\prime,General,v,\tau,1)$ to all;\\

{\bf at any time: }\\
\tb \textbf{if} received $(echo^\prime,General,v,\tau,1)$ from $\ge
n-2f$ distinct nodes
\textbf{then}\\
\tre send $(echo^\prime,General,v,\tau,1)$ to all;\\
\tb \textbf{if} received $(echo^\prime,General,v,\tau,1)$ from $\ge
n-f$ distinct nodes
\textbf{then}\\
\tre accept $(General,v,\tau,1)$;
\normalsize

\end{minipage} }
\caption{\ConsBCast} \label{alg:init-bdcst}
\end{figure}

We first prove the properties of the \ConsBCast primitive and later
we prove the correctness of the \ByzConsensus\ procedure.\\ The
$\ConsBCast$ primitive and the \Broadcast primitive (defined in
Section~\ref{sec:bcast-primitive}) satisfy the following [TPS-*]
properties of Toueg, Perry and Srikanth~\cite{FastAgree87}, which
are phrased in our system model.
\begin{enumerate}
\item[\bf TPS-1] \emph{(Correctness)} If a correct node $p$ does \Broadcast
$(p,m,\tau,k)$ by $\tau+(2k-2)\bar d$ on its timer, then every
correct node accepts $(p,m,\tau,k)$ by $\tau+2k\bar d$ on its timer.
  \item[\bf TPS-2] {\em (Unforgeability)} If no correct node $p$
  does a
\Broadcast $(p,m,\tau,k),$ then no correct node accepts
$(p,m,\tau,k).$
  \item[\bf TPS-3] {\em (Relay)} If a correct node accepts $(p,m,\tau,k)$ by
$\tau+2r\bar d,$ for $r\ge k,$ on its timer then every other correct
node accepts $(p,m,\tau,k)$ by $\tau+(2r+2)\bar d$ on its timer.
  \item[\bf TPS-4] {\em (Detection of broadcasters)} If a correct node
accepts $(p,m,\tau,k)$ by $\tau+2r\bar d,$ on its timer then every
correct node has $p\in broadcasters$ by $\tau+(2k+1)\bar d$ on its
timer. Furthermore, if a correct node $p$\, does not \Broadcast any
message, then a correct node can never have $p\in broadcasters.$
\end{enumerate}

Additionally, the \ConsBCast primitive also satisfies:
\begin{enumerate}
\item[\bf TPS-5] {\em (Uniqueness)} If a correct node accepts $(General,m,\tau,1),$
then no correct node ever accepts $(General,m^\prime,\tau,1)$ with
$m^\prime\neq m.$
\end{enumerate}

Notice the differences from the original properties. The detection
property does not require having $r\ge k.$ In general, the relay
property holds even earlier than $r\ge k.$ The condition $r\ge k$ of
when the property can be guaranteed is used to simplify the possible
cases.  At $r<k,$ if an accept takes place as a result of getting
$n-f$ echo messages, the adversary may cause the relay to take
$3\bar d$\ by rushing messages to one correct node and delay
messages to and from others.

\begin{theorem} The \ConsBCast primitive satisfies the five [TPS-*]
properties.
\end{theorem}

\begin{proof}
\mbox{\ }\\
\noindent \emph{Correctness}: If all correct nodes send
$(echo,General,v,\tau,1)$ at time $\tau$ on their timers, then by
Lemma~\ref{lemma-1phase} every correct node accepts
$(General,v,\tau,1)$ from $n-f$ correct nodes by $\tau+\bar d$ on
its timer. Thus each correct node sends $(echo,General,v,\tau,1)$ by
that time and will accept $(General,v,\tau,1)$ by
$\tau+2\bar d$ on their timers.\\

\noindent{\em Unforgeability}: If all correct nodes hold the same
initial value $v$ then no  correct node will send
$(echo,General,v^\prime,1),$ thus no correct node will receive $n-f$
distinct $(echo,General,v^\prime,1)$ messages. Therefore, no correct
node will send $(echo^\prime,General,v^\prime,1),$ and no correct
node will ever receive $n-2f$ or $n-f$ distinct
$(echo^\prime,General,v^\prime,1)$
messages.  Thus, no correct node can accept $(General,v^\prime,1).$\\

\noindent{\em Relay}: If a correct node accepts $(General,v,\tau,1)$
by $\tau+2r\bar d$ on its timer, then it received $n-f$ distinct
$(echo^\prime,General,v,\tau,1)$ message by that time. $n-2f$ of
these were sent by correct nodes and by Lemma~\ref{lemma-1phase} all
of them will reach all correct nodes by $\tau+(2r+1)\bar d.$ As a
result, all such correct nodes will send
$(echo^\prime,General,v,\tau,1),$ which will be received by all
correct nodes.  Hence, by $\tau+(2r+2)\bar d$ on their timers, all
correct nodes will hold $n-f$ distinct
$(echo^\prime,General,v,\tau,1)$ messages and will thus accept
$(General,v,\tau,1).$\\

\noindent{\em Detection of broadcasters}: If a correct node
$q^\prime$ accepts $(General,v,\tau,1)$ by time $\tau+2r\bar d$ on
its timer, then node $q^\prime$ should have received at least $n-f$
distinct $(echo^\prime,General,v,\tau,1)$ messages, at least $n-2f$
of which are from correct nodes. Let $q$ be the first correct node
to ever send $(echo^\prime,General,v,\tau,1).$  If $q$ sent it as a
result of receiving $n-f$ such messages, then $q$ is not the first
to send. Therefore, it should have sent it as a result of receiving
$n-f$ $(echo,General,v,\tau,1)$ messages by time $\tau+\bar d.$
Thus, at least $n-2f$ such messages were sent by correct nodes by
time $\tau$ on their timers and would arrive at all correct nodes by
time $\tau+\bar d$ on their timers.  As a result, all will have
$General\in broadcasters.$\\

\noindent\emph{Uniqueness}: Notice that if a correct node sends
$(echo^\prime,General,v,\tau,1)$ by time $\tau+\bar d,$ then no
correct node sends $(echo^\prime,General,v^\prime,1)$ at any later
time. Otherwise, similarly to the arguments in proving the previous
property we get that at least $n-f$ nodes sent
$(echo,General,v,\tau,1)$ and $n-f$ nodes sent
$(echo,General,v^\prime,1).$ Since $n>3f,$ this implies that at
least one correct node sent both $(echo,General,v,\tau,1)$ and
$(echo,General,v^\prime,1),$ and this is not allowed.\\

Also note that if a correct node accepts $(General,v,\tau,1),$ then
at least one correct node sends $(echo^\prime,General,v,\tau,1),$
which yields the proof of the \emph{Uniqueness} property. \qed
\end{proof}

Nodes stop participating in \ByzConsensus\ when they are instructed
to do so. They stop participating in the \Broadcast primitive $2\bar
d$ after they terminate \ByzConsensus.

\begin{definition}
\item A node {\bf returned} a value $m$ if it has stopped
and returned $value=m.$

\item A node $p$ {\bf decides} if it
stops at that timer time and  returns a $value \ne \perp.$

\item A node $p$ {\bf aborts} if it stops and returns
$\perp.$

\end{definition}

\begin{theorem}\label{thm:byz}
The \ByzConsensus\ procedure satisfies the  Termination property.
When $n>3f,$ it also satisfies Agreement, Validity and the two early
stopping conditions.
\end{theorem}

\begin{proof} We prove the five properties of the theorem.
We build up the proof through the following arguments.

\begin{lemma}\label{lemma2} If a correct node aborts at time
$T+2r\bar d$ on its timer, then no correct node decides at a time
$T+2r^\prime\bar d \ge T+2r\bar d$ on its timer.
\end{lemma}

\begin{proof} Let $p$ be a correct node that aborts at time $T+2r\bar d.$ In this
case it should have identified exactly $r-2$ broadcasters by that
time.  By the detection of broadcasters property [TPS-4] no correct
node will ever accept $(General,v,T,1)$ and $r-2$ distinct messages
$(q_i,v,T,i)$ for $2\le i\le r-1,$ since that would have caused all
correct nodes to hold $r-1$ broadcasters by time $T+(2r-1)\bar d$ on
their timers. Thus, no correct node can decide at local-time
$T+2r^\prime\bar d \ge T+2r\bar d.$ \qed

\begin{lemma}\label{lemma1} If a correct node decides by time
$T+2r\bar d$ on its timer, then every correct node decides by time
$T+2(r+1)\bar d$ on its timer.
\end{lemma}

\begin{proof}

Let $p$ be a correct node that decides by time $T+2r\bar d$ on its
timer. We consider the following cases:

\begin{enumerate}

\item $ r=1:$ No correct node can abort by time $T+2\bar d,$ since the
inequality will not hold. Node $p$ must have accepted
$(General,v,T,1)$ by $T+2\bar d.$ By the relay property [TPS-3] all
correct nodes will accept $(General,v,T,1)$ by $T+4\bar d$ on their
timers. Moreover, $p$ invokes {\bf \Broadcast}$(p,v,T,2),$ by which
the correctness property [TPS-1] will be accepted by all correct
nodes by time $T+4\bar d$ on their timers. Thus, all correct nodes
will have $value\ne \perp$ and will \Broadcast and stop by time
$T+4\bar d$ on their timers.

\item $2 \le r \le f+1.$ Node $p$ must have accepted $(General,v,T,1)$
and also accepted $r-1$ distinct $(q_i,v,T,i)$ messages for all $i,
2 \le i \le r,$ by time $T+2r\bar d$ on its timer. By
Lemma~\ref{lemma2}, no correct aborts by that time. By Relay
property [TPS-3] each $(q_i,v,T,i)$ message will be accepted by all
correct nodes by time $T+(2r+2)\bar d$ on their timers. Node $p$
does \Broadcast $(p,v,T,r+1)$ before stopping. By the correctness
property, this message will be accepted by all correct nodes by time
$T+(2r+2)\bar d$ on their timers. Thus, no correct node will abort
by $T+(2r+2)\bar d$ and all correct nodes will have $value\ne \perp$
and will decide and stop by that time.

\item $r=f+2.$ Node $p$ must have accepted $(q_i,v,T,i)$ messages
for all $i, 2 \le i \le f+2,$ by $T+(2f+4)\bar d$ on its timer,
where the $f+1$ $q_i$'s are distinct. At least one of these $f+1$
nodes, say $q_j,$ must be correct. By the Unforgeability property
[TPS-2] $q_j,$ invoked \Broadcast $(q_j,v,T,j)$ by time $T+(2j)\bar
d$ on its timer, and decided. Since $j \le f+1$ the above arguments
imply that by $T+(2f+4)\bar d$ on their timers all correct will
decide.
\end{enumerate}
\qed

\end{proof}

\noindent Lemma~\ref{lemma1} implies that if a correct node decides
at time $T+2r\bar d$ on its timer, then no correct node aborts at
round $T+2r^\prime\bar d.$ Lemma~\ref{lemma2} implies the
other direction.\\

\noindent{\textbf{Termination}:} Lemma~\ref{lemma1} implies that if
any correct node decides, all decide and stop.  Assume that no
correct node decides.  In this case, no correct node ever invokes a
\Broadcast $(q,v,T,\_).$  By detection of broadcasters property
[TPS-4], no correct node will ever be considered as broadcaster.
Therefore, by time $T+((2f+4)\bar d$ on their timers, all correct
nodes will have at most $f$ broadcasters and will abort and
stop.\qed
\mbox{\ }\\

\vspace{-3mm}\noindent{\textbf{Agreement}:} If no correct node
decides, then all abort, and return to the same value. Otherwise,
let $p$ be the first correct node to decide. Therefore, no correct
node aborts. The value returned by $p$ is the value $v$ of the
accepted $(General,v,1)$ message. By Properties [TPS-3] and [TPS-5]
all correct nodes accept $(General,v,T,1)$ and no correct node
accepts $(General,v^\prime,T,1)$ for $v \ne v^\prime.$ Thus all
correct nodes return the same value.\qed
\mbox{\ }\\

\vspace{-3mm} \noindent{\textbf{Validity}:} Let all the correct
nodes begin with the same value $v^\prime$ and invoke the protocol
with the same timer time ($T$). Then, by time $T+\bar d$ on their
timers, all correct nodes receive at least $n-2f$ distinct
$(echo,General,v^\prime,T,1)$ messages via  the \ConsBCast primitive
and send $(echo^\prime,General,v^\prime,T,1)$ messages to all.
Hence, all nodes receive at least $n-f$ distinct
$(echo^\prime,General,v^\prime,T,1)$ messages by $T+2\bar d$ on
their timers and thus accept $(General,v^\prime,T,1).$ Hence in the
\ByzConsensus\ procedure all correct nodes set their value to
$v^\prime.$ By $T+2\bar d$ on their timers,  all correct nodes will
stop and return $v^\prime.$\qed
\mbox{\ }\\

\vspace{-3mm}\noindent{\textbf{Early-stopping}:} The first early
stopping property [ES-1] is directly implied from the proof of the
validity property. Correct nodes proceed once they receive messages
from $n-f$ nodes, thus it is enough to receive messages from all
correct nodes. The proof of the second early stopping property
[ES-2] is identical to the proof of the termination property.  By
time $T+(2f^\prime+4)\bar d$ all will abort unless any correct node
invokes \Broadcast by that time on its timer. This implies that by
$T+(2f^\prime+6)\bar d$ on their timers all correct nodes will
always terminate, if the actual number of faults $f^\prime$ is less
than $f.$\qed \mbox{\ }

Thus the proof of the theorem is concluded.\qed

\end{proof}

\subsection{The \Broadcast Primitive}\label{sec:bcast-primitive} This
section presents the \textbf{Broadcast} (and \emph{accept})
primitive that is used by the \ByzConsensus\ procedure presented
earlier, in Section~\ref{sec:byzssconsensus}. The primitive follows
the primitive of of Toueg, Perry, and Srikanth~\cite{FastAgree87},
though here it is presented in a real-time model.

In the original synchronous model, nodes advance according to
phases. This intuitive lock-step process clarifies the presentation
and simplifies the proofs.  In this section, the discussion
carefully considers the various time consideration and proves that
nodes can rush through the protocol and do not to need to wait for a
completion of a ``phase'' in order to move to the next step of the
protocol.

Note that when a node invokes the procedure it evaluates all the
messages in its buffer that are relevant to the procedure.

\begin{figure}[!h]
\center \fbox{\begin{minipage}{4.5in} \footnotesize
\setlength{\baselineskip}{3.5mm}

\noindent Procedure \textbf{\Broadcast}$(p,m,\tau,k)$\\
\mbox{\ }\hfill\textit{/* executed per such quadruple */}\\
\mbox{\ }\hfill\textit{/* nodes send specific message with the same $\tau$ only once */}\\
\mbox{\ }\hfill\textit{/* multiple  messages
 sent by an  individual node are ignored */}\\

node $p$ sends $(init,p,m,\tau,k)$ to all nodes;\\

{\bf by time }$(\tau+(2k-1)\bar d):$ \\
\tb \textbf{if} (received $(init,p,m,\tau,k)$ from $p$  \textbf{then}\\
\tre send $(echo,p,m,\tau,k)$ to all;\\

{\bf by time }$(\tau+2k\bar d):$ \\
\tb \textbf{if} (received $(echo,p,m,\tau,k)$ from $\ge n-2f$
distinct nodes $q$ \textbf{then}\\
\tre send $(init^\prime,p,m,\tau,k)$ to all;\\
\tb \textbf{if} (received $(echo,p,m,\tau,k)$ msgs from $\ge n-f$
distinct nodes \textbf{then}\\
\tre  accept $(p,m,\tau,k)$;\\

{\bf by time }$(\tau+(2k+1)\bar d):$ \\
\tb \textbf{if} (received $(init^\prime,p,m,\tau,k)$
from $\ge n-2f$ \textbf{then}\\
\tre  $broadcasters:=broadcasters\bigcup\{p\}$;\\
\tb \textbf{if} (received $(init^\prime,p,m,\tau,k)$ from $\ge n-f$
distinct nodes \textbf{then}\\
\tre  send $(echo^\prime,p,m,\tau,k)$ to all;\\

{\bf at any time: } \\
\tb \textbf{if} (received $(echo^\prime,p,m,\tau,k)$ from $\ge
n-2f$ distinct nodes \textbf{then}\\
\tre  send $(echo^\prime,p,m,\tau,k)$ to all;\\
\tb \textbf{if} (received $(echo^\prime,p,m,\tau,k)$ from $\ge n-f$
distinct nodes) \textbf{then}\\
\tre  accept $(p,m,\tau,k)$;\\
\textbf{end}
\normalsize
\end{minipage} }
\caption{\Broadcast primitive} \label{alg:reg-bdcst}
\end{figure}

The \Broadcast primitive satisfies the four [TPS-*] properties,
under the assumption that $n>3f.$ The proofs below follow closely to
the original proofs of~\cite{FastAgree87}, in order to make it
easier for readers that are familiar with the original proofs.

\begin{lemma}\label{lemma-1phase}
If a correct node $p_i$ sends a message at timer time $T_i\le
\tau+r\bar d$ on $p_i$'s timer it will be recieved by each correct
node $p_j$ by timer time $\tau+(r+1)\bar d$ on $p_j$'s timer.
\end{lemma}

\begin{proof}
Assume that node $p_i$ sends a message at real time $t$ with timer
time $T_i(t)\le \tau+r\bar d.$ Thus,
$T_i(t)\le\tau+r(\bsigma+d)(1+\rho).$ It should arrive at every
correct timer $p_j$ within $d(1+\rho)$ on any correct node's timer.
Recall that $|T_i(t)-T_j(t)|<\bsigma(1+\rho).$ If $T_j\ge T_i$ we
are done. Otherwise, $$T_j(t)\le T_i(t) + \bsigma(1+\rho)\le
\tau+r(\bsigma+d)(1+\rho)+\bsigma(1+\rho)\;.$$ By the time (say
$t^\prime$) that the message arrives to $p_j$ we get
$$T_j(t^\prime)\le
\tau+r(\bsigma+d)(1+\rho)+\bsigma(1+\rho)+d(1+\rho)\le
\tau+(r+1)\bar d\;.$$ \mbox{\ }\qed
\end{proof}

\begin{lemma}\label{lemma4.1}
If a correct node ever sends $(echo^\prime,p,m,\tau,k)$ then at
least one correct node must have sent $(echo^\prime,p,m,\tau,k)$ by
timer time $\tau+(2k+1)\bar d.$
\end{lemma}

\begin{proof}
Let $t$ be the earliest timer time by which any correct node $q$
sends the message $(echo^\prime,p,m,\tau,k).$ If $t> \tau+(2k+1)\bar
d,$ node $q$ should have received $(echo^\prime,p,m,\tau,k)$ from
$n-2f$ distinct nodes, at least one of which from a correct node
that was sent prior to timer time $\tau+(2k+1)\bar d.$ \mbox{\ }\qed
\end{proof}

\begin{lemma}\label{lemma4.2}
If a correct node ever sends $(echo^\prime,p,m,\tau,k)$ then $p$'s
$(init,p,m,\tau,k)$ must have been received by at least one correct
node by time $\tau+(2k-1)\bar d.$
\end{lemma}

\begin{proof}
By Lemma~\ref{lemma4.1}, if a correct node ever sends
$(echo^\prime,p,m,\tau,k),$ then some correct node $q$ should send
it by time timer $\tau+(2k+1)\bar d.$ By the procedure, $q$ have
received $(init^\prime,p,m,\tau,k)$ from at least $n-f$ nodes by
timer time $\tau+(2k+1)\bar d.$ At least one of them is correct who
have received $n-2f$ $(echo,p,m,\tau,k)$ by timer time $\tau+2k\bar
d.$ One of which was sent by correct node that should have received
$(init,p,m,\tau,k)$ before sending $(echo,p,m,\tau,k)$ by timer time
$\tau+(2k-1)\bar d.$ \mbox{\ }\qed
\end{proof}

\begin{theorem}\label{thm4.1}
The \Broadcast primitive presented in Figure~\ref{alg:reg-bdcst}
satisfies properties [TPS-1] through [TPS-4].
\end{theorem}

\begin{proof}
\mbox{\ }

{\bf Correctness:} Assume that a correct node $p$ sends
$(p,m,\tau,k)$ by $\tau+(2k-2)\bar d$ on its timer. Every correct
node receives $(init,p,m,\tau,k)$ and sends $(echo,p,m,\tau,k)$ by
$\tau+(2k-1)\bar d$ on its timer. Thus, every correct node receives
$n-f$ $(echo,p,m,\tau,k)$ from distinct nodes by $\tau+(2k-1)\bar d$
on its timer and accepts $(p,m,\tau,k).$

{\bf Unforgeability:} If no correct node $p$ does a \Broadcast
$(p,m,\tau,k),$ it does not send $(init,p,m,\tau,k),$ and no correct
node will send $(echo,p,m,\tau,k)$ by $\tau+(2k-1)\bar d$ on its
timer. Thus, no correct node accepts $(p,m,\tau,k)$ by $\tau+2k\bar
d$ on its timer. If a correct node would have accepted
$(p,m,\tau,k)$ at a later time it can be only as a result of
receiving $n-f$ $(echo^\prime,p,m,\tau,k)$ distinct messages, some
of which must be from correct nodes. By Lemma~\ref{lemma4.2}, $p$
should have sent $(init,p,m,\tau,k),$ a contradiction.

{\bf Relay:} Notice that $r\ge k,$ thus even if nodes issue an
accept at earlier time, the claim holds for the specified times.

The subtle point is when a correct node issues an accept as a result
of getting echo messages. If $r=k$ and the correct node, say $q,$
have received $(echo,p,m,\tau,k)$ from $n-f$ nodes by $\tau+2k\bar
d$ on its timer. At least $n-2f$ of them were sent by correct nodes.
Since every correct node among these has sent its message by
$\tau+(2k-1)\bar d,$ all those messages should have arrived to every
correct node by $\tau+2k\bar d$ on its timer. Thus, every correct
node should have sent $(init^\prime,p,m,\tau,k)$ by $\tau+2k\bar d$
on its timer. As a result, every correct node will receive $n-f$
such messages by $\tau+(2k+1)\bar d$ on its timer and will send
$(echo^\prime,p,m,\tau,k)$ by that time, which will lead all correct
nodes to accept $(p,m,\tau,k)$ by $\tau+(2r+2)\bar d$ on its timer.

Otherwise, the correct node, say $q,$ accepts $(p,m,\tau,k)$ by
$\tau+2r\bar d$ on its timer as a result of receiving $n-f$
$(echo^\prime,p,m,\tau,k)$ by that time.  Since $n-f$ of these are
from correct nodes, they should arrive at any correct node by
$\tau+(2r+1)\bar d$ on their timers. As a result, by
$\tau+(2r+1)\bar d,$ all correct nodes would send
$(echo^\prime,p,m,\tau,k)$ and by $\tau+(2r+2)\bar d$ on their
timers all will accept $(p,m,\tau,k).$

 {\bf Detection of
broadcasters:} As in the original proof, we first argue the second
part.  Assume that a correct node $q$ adds node $p$ to
$broadcasters.$ It should have received $n-2f$
$(init^\prime,p,m,\tau,k)$ messages.  Thus, at least one correct
node has sent $(init^\prime,p,m,\tau,k)$ as a result of receiving
$n-2f$ $(echo,p,m,\tau,k)$ messages. One of these should be from a
correct node that has received the original \Broadcast message of
$p.$

To prove the first part, we consider two similar cases to support
the Relay property. If $r=k$ and the correct node, say $q,$ accepts
$(p,m,\tau,k)$ as a result of receiving $n-f$ $(echo,p,m,\tau,k)$ by
$\tau+2k\bar d$ on its timer. At least $n-2f$ of them were sent by
correct nodes. Since every correct node among these has sent its
message by $\tau+(2k-1)\bar d,$ all those messages should have
arrived at every correct node by $\tau+2k\bar d$ on its timer. Thus,
every correct node should have sent $(init^\prime,p,m,\tau,k)$ by
$\tau+2k\bar d$ on its timer. Consequently, all correct nodes will
receive $n-f$ such messages by time $\tau+(2k+1)\bar d$ and will add
$p$ to $broadcasters.$

Otherwise, $q$ accepts $(p,m,\tau,k)$ as a result of receiving
$(echo^\prime,p,m,\tau,k)$ from $n-f$ nodes by $\tau+2r\bar d$ (for
$r\ge k$) on its timer. By Lemma~\ref{lemma4.1} a correct node sent
$(echo^\prime,p,m,\tau,k)$ by $\tau+(2k+1)\bar d.$ It should have
received $n-f$ $(init^\prime,p,m,\tau,k)$ messages by that time. All
such messages that were sent by correct nodes were sent by
$\tau+2k\bar d$ on their timers and should arrive at every correct
node by $\tau+(2k+1)\bar d$ on its timer. Since there are at least
$n-2f$ such messages, all will add $p$ to $broadcasters$ by
$\tau+(2k+1)\bar d$ on their timers. \mbox{\ }\qed

\end{proof}

\end{proof}


\begin{thebibliography}{XX}
\small

\vspace{-2mm}
\bibitem{CSEVAL98}E. Anceaume, I. Puaut, ``{\em Performance Evaluation
of Clock Synchronization Algorithms}'',
 Technical report 3526,INRIA, 1998.

\ignore{-2mm}
\bibitem{ADG91}A. Arora, S. Dolev, and M.G. Gouda, ``{\em Maintaining digital clocks
in step}'', Parallel Processing Letters, 1:11-18, 1991.

\ignore{
\ignore{-2mm}
\bibitem{AKMSV}B. Awerbuch, S. Kutten, Y. Mansour, B. Patt-Shamir and G.
Varghese, ``{\em Time Optimal Self-Stabilizing Synchronization},
Proceedings of the 25th Symp. on Theory of Computing, 1993.
}


\ignore{-2mm}
\bibitem{R27}J. Brzezi\`{n}ski, and M. Szychowiak, ``{\em Self-Stabilization in
Distributed Systems - a Short Survey}, Foundations of Computing
and Decision Sciences, Vol. 25, no. 1, 2000.

\ignore{
\ignore{-2mm}
\bibitem{Prob-clock}
C. Cristian, ``{\em Probabilistic Clock Synchronization}'',
Distributed Computing, vol. 3 pp. 146-158, 1989. }

\ignore{ \ignore{-2mm}
\bibitem{pulse-tr}
A. Daliot, D. Dolev, H. Parnas, ``{\em Self-Stabilizing Pulse
Synchronization Inspired by Biological Pacemaker Networks}", {\bf
Technical Report TR2003-1}, Schools of Engineering and Computer
Science, The Hebrew University of Jerusalem, March
2003.\\
Url:
http://leibnitz.cs.huji.ac.il/tr/acc/2003/HUJI-CE-LTR-2003-1\_pulse-tr6.ps
}

\ignore{-2mm}
\bibitem{bio-pulse-synch}
A. Daliot, D. Dolev and H. Parnas, ``{\em Self-stabilizing Pulse
Synchronization Inspired by Biological Pacemaker Networks}'', Proc.
of the 6th Symposium on Self-stabilizing Systems (SSS'03
San-Francisco), pp. 32-48, 2003.

\bibitem{NEW-PULSE-TR}
A. Daliot and D. Dolev, ``{\em Self-stabilizing Byzantine Pulse
Synchronization }", {Technical Report TR2005-84}, Schools of
Engineering and Computer Science, The Hebrew University of
Jerusalem, August 2005. A revised version appears in
http://arxiv.org/abs/cs.DC/0608092 .

\ignore{-2mm}
\bibitem{R12}
D. Dolev, J. Halpern, and H. R. Strong, ``{\em On the Possibility
and Impossibility of Achieving Clock Synchronization}'', J. of
Computer and Systems Science, Vol. 32:2, pp. 230-250, 1986.

\ignore{-2mm}
\bibitem{PolyAgree82}
D. Dolev, H. R. Strong, ``{\em Polynomial Algorithms for Multiple
Processor Agreement}'', In Proceedings, the 14th ACM SIGACT
Symposium on Theory of Computing, 401-407, May 1982. (STOC-82)

\ignore{-2mm}
\bibitem{R3}
D. Dolev, J. Y. Halpern, B. Simons, and R. Strong, ``{\em Dynamic
Fault-Tolerant Clock Synchronization}'', J. Assoc. Computing
Machinery, Vol. 42, No.1, pp. 143-185, Jan. 1995.

\ignore{-2mm}
\bibitem{Rapprox}D. Dolev, N. A. Lynch, E. Stark, W. E. Weihl and S.
Pinter, ``{\em Reaching Approximate Agreement in the Presence of
Faults}'',  J. of the ACM, 33 (1986) 499-516.


\ignore{-2mm}
\bibitem{DPossImp97}S. Dolev, ``{\em Possible and Impossible Self-Stabilizing Digital Clock
Synchronization in General Graphs}'', Journal of Real-Time
Systems, no. 12(1), pp. 95-107, 1997.


\ignore{-2mm}
\bibitem{R22}
S. Dolev, ``{\em Self-Stabilization}'', The MIT Press, 2000.

\ignore{ \ignore{-2mm}
\bibitem{DolSSConsensus}
S. Dolev, S. Rajsbaum, ``{\em Stability of Long-lived
Consensus}'', Proc. of the 19th Annual ACM Symp. on Principles of
Distributed Computing, 2000.
}

\ignore{-2mm}
\bibitem{DolWelSSBYZCS04}
S. Dolev, and J. L. Welch, ``{\em Self-Stabilizing Clock
Synchronization in the presence of Byzantine faults}'', Journal of
the ACM, Vol. 51, Issue 5, pp. 780 - 799, 2004.

\ignore{-2mm}
\bibitem{DW97b}S. Dolev and J. L. Welch, ``{\em Wait-free clock
synchronization}'', Algorithmica, 18(4):486-511, 1997.

\ignore{
\ignore{-2mm}
\bibitem{FetCri97}
C. Fetzer and F. Cristian, ``{\em Integrating External and
Internal Clock Synchronization}'', Real-Time Systems, vol 12,
1997, pp. 123-171.
}

\ignore{
\ignore{-2mm}
\bibitem{Fetzer:1995}
C. Fetzer and F. Cristian, ``{\em An Optimal Internal Clock
Synchronization Algorithm}'', Proceedings of the $10^{th}$
Conference on Computer Assurance, 1995, pp. 187-196, Gaithersburg,
MD, USA.
}

\ignore{-2mm}
\bibitem{Impossibility86}
M. J. Fischer, N. A. Lynch and M. Merritt, ``{\em Easy
impossibility proofs for distributed consensus problems}'',
Distributed Computing, Vol. 1, pp. 26-39, 1986.

\ignore{-2mm}
\bibitem{H00b}T. Herman, ``{\em Phase clocks for transient fault repair}'', IEEE
Transactions on Parallel and Distributed Systems,
11(10):1048-1057, 2000.

\ignore{ \ignore{-2mm}
\bibitem{Lamport:1985:SCP}
L. Lamport and P. M. {Melliar-Smith}, ``{\em Synchronizing Clocks
in the Presence of Faults}'', Journal of the ACM, vol. 32(1), pp.
52-78, 1985. }


\ignore{-2mm}
\bibitem{Liskov:1991}
B. Liskov, ``{\em Practical Use of Synchronized Clocks in
Distributed Systems}'', Proceedings of $10^{th}$ ACM Symposium on
the Principles of Distributed Computing, 1991, pp. 1-9.

\ignore{ \ignore{-2mm}
\bibitem{R21}
J. Lundelius, and N. Lynch, ``{\em An Upper and Lower Bound for
Clock Synchronization},'' Information and Control, Vol. 62, pp.
190-205, Aug/Sep. 1984.
}

\ignore{ \ignore{-2mm}
\bibitem{R9}
N. Lynch, ``{\em Distributed Algorithms}'', Morgan Kaufmann, 1996.
}

\ignore{-2mm}
\bibitem{shamir-clcks}
B. Patt-Shamir, ``{\em A Theory of Clock Synchronization}'',
Doctoral thesis, MIT, Oct. 1994.

\ignore{-2mm}
\bibitem{PT97}M. Papatriantafilou, P. Tsigas,
``{\em On Self-Stabilizing Wait-Free Clock Synchronization}'',
Parallel Processing Letters, 7(3), pages 321-328, 1997.

\ignore{
 \ignore{-2mm}
\bibitem{Paulitsch:2002}
M. Paulitsch, ``{\em Fault-Tolerant Clock Synchronization for
Embedded Distributed Multi-Cluster Systems}'', Doctoral thesis,
Treitlstr. 1-3/3/182-1, Vienna, Austria, Institut f{\``u}r
Technische Informatik, Technische Universit{\''a}t Wien, 2002.
}

\ignore{-2mm}
\bibitem{Schneider87}
F. Schneider, ``{\em Understanding Protocols for Byzantine Clock
Synchronization}'', Technical Report 87-859, Dept. of Computer
Science, Cornell University, Aug. 1987.

\ignore{-2mm}\bibitem{FastAgree87} S. Toueg, K. J. Perry, T. K.
Srikanth, ``{\em Fast Distributed Agreement}'', SIAM Journal on
Computing, 16(3):445-457, June 1987.

\ignore{-2mm}
\bibitem{WelchLynch88}
J. L. Welch, and N. Lynch, ``{\em A New Fault-Tolerant Algorithm
for Clock Synchronization}'', Information and Computation 77,
1-36, 1988.

\end{thebibliography}
\end{document}